\documentclass[12pt]{article}
\usepackage{amssymb,amsmath,epsfig}

\begin{document}

\title{\bf Dissipative Spherical Collapse of Charged Anisotropic Fluid in $f(R)$ Gravity}

\author{H. Rizwana Kausar\thanks{rizwa\_math@yahoo.com}\\
Director, Centre for Applicable Mathematics \& Statistics,\\
University of Central Punjab, Lahore, Pakistan. \\ Ifra Noureen \thanks{ifra.noureen@gmail.com}\\
University of Management and Technology, Lahore, Pakistan.}
\date{}
\maketitle

\begin{abstract}
This manuscript is devoted to study the combined effect of a viable
$f(R)=R+{\alpha}{R^n}$ model and electromagnetic field on the
instability range of gravitational collapse. We assume charged
anisotropic fluid that dissipate energy via heat flow and discuss
that electromagnetic field, density inhomogeneity, shear and phase
transition on astrophysical bodies can be incorporated by locally
anisotropic background. Dynamical equations help to investigate the
evolution of self-gravitating objects and leads to the conclusion
that adiabatic index depend upon the electromagnetic background,
mass and radius of the spherical objects.
\end{abstract}

\section{Introduction}

Gravitational collapse is highly dissipative phenomenon. The effects
of dissipation describe a wide range of situations. For example,
using quasi-static approximation, limiting cases of radiative
transport have been studied in \cite{52}. It is found that
hydrostatic time scale is very small as compared to the stellar
lifetimes for different phases of a star's life. It is of the order
of 27 minutes for the sun, 4.5 seconds for a white dwarf and
$10^{-4}$ seconds for a neutron star of one solar mass and 10 km
radius \cite{53}. The dissipative factors enhances instability range
at Newtonian limits but develop more stability at relativistic
annexes. The impressions of radiation, anisotropy and shearing
viscosity at Newtonian and post-Newtonian eras are inquired in
\cite{5}-\cite{8}. Various prospects of collapse phenomenon in
account with dark source are worked out in recent past
\cite{Riz1}-\cite{Riz6}. Due to high dissipation, matter produce
large amount of charge in collapsing phenomenon and so it is well
motivated problem to investigate electromagnetic field effects on
the gravitational collapse \cite{V}.

Chandrasekhar \cite{1} took initiative to workout dynamical
instability problem. Dynamical instability is pragmatic in
establishing the evolution and formation of stellar objects that
must be stable against fluctuations. Generally, adiabatic index
$\Gamma$ is utile to address instability problem. Isotropic spheres
of mass $M$ and $R$ radius may be related as
$\Gamma\geq\frac{4}{3}+n\frac{M}{r}$, where number $n$ depends upon
the star's structure. Later on instability range for anisotropic,
adiabatic, non-adiabatic and shearing viscous fluids has been
examined in \cite{2}-\cite{4}. Besides $\Gamma$ many other matter
variables such as dissipation, radiation, shearing stress,
anisotropy, expansion-free condition, etc. may also be responsible
for dynamical instability and evolution in stars depending upon
fluid properties.

Modified theories of gravity have received enormous attention in
recent years. Inclusion of higher order curvature invariants and
coupled scalar fields have become a paradigm in alternative gravity
theories. For this purpose, various alterations are made in
Einstein-Hilbert (EH) action \cite{TP}-\cite{Sn}. The elementary and
likely modification is to include curvature terms that are of type
$f(R)$ having combinations of Ricci scalar $R$. In this way, gravity
tends to modify on large scales that reveals enormous observational
signatures like modified galaxy clustering spectrum \cite{cb1, cb2},
weak lensing \cite{st1, st2} and cosmic microwave background
\cite{zs1, zs2}.

The most studied and simplest models in $f(R)$ theory are
$f(R)=R+\sigma\frac{\mu^4}{R}$ and $f(R)=R+\alpha R^2$, where
$\sigma=\pm1$, $\alpha$ is a positive real number and $\mu$ is a
parameter with units of mass. Usually, the positive values of scalar
curvature depicts standard cosmological corrections leading to
de-Sitter space \cite{Starobinsky} whereas negative values help to
discuss accelerating universe due to dark energy \cite{CDTT}. The
effects of these $f(R)$ models on the dynamical instability of
gravitational collapse has been discussed in recent papers
\cite{Riz10, Riz11}. In the same context, Sharif and Yousaf
established the range of instability for charged expansion-free,
dissipative collapse for spherical and cylindrical symmetries in
$f(R)$ gravity \cite{Z1}-\cite{Z3}. In this paper, we adopted
$f(R)=R+\alpha R^n$ to discuss dynamical instability of
gravitational collapse in the background of electromagnetic field.

The manuscript is arranged as follows. In section \textbf{2},
energy-momentum tensor of matter distribution along with Maxwell's
and Einstein's field equations is given. Section \textbf{3} provide
the knowledge about adopted $f(R)$ model and perturbation scheme. In
the same section, instability range would be discussed for Newtonian
and post-Newtonian regimes in the form of $\Gamma$. Final section
\textbf{4} provides summary of the paper and followed by an
appendix.

\section{Evolution Equations}

We have chosen timelike three dimensional spherical boundary surface $\Sigma$ that
delimitates four dimensional line element into two realms termed as exterior and
interior region. Interior region inside the boundary is
\begin{equation}\label{1}
ds^2_-=A^2(t,r)dt^{2}-B^2(t,r)dr^{2}-C^2(t,r)(d\theta^{2}+\sin^{2}\theta d\phi^{2}).
\end{equation}
while line element for exterior region \cite{Z3} is considered as
\begin{equation}\label{25}
ds^2_+=\left(1-\frac{2M}{r}+\frac{Q^2}{r^2}\right)d\nu^2+2drd\nu-r^2(d\theta^2+\sin^{2}\theta
d\phi^{2}).
\end{equation}
Here $\nu$ corresponds to retarded time, $M$ is the total mass and
$Q$ indicates the total charge of fluid.

The generalized EH action for $f(R)$ gravity in account with Maxwell source modifies to
\begin{equation}\label{b}
S=\frac{1}{2}\int
d^{4}x\sqrt{-g}\left(\frac{f(R)}{\kappa}-\frac{\digamma}{2\pi}\right).
\end{equation}
Here $\kappa$ stands for coupling constant and
$\digamma=\frac{1}{4}F^{uv}F_{uv}$ is the Maxwell invariant. We use
metric approach to recover field equations by varying above action
with $g_{uv}$ as follows
\begin{equation}\label{b'}
f_RR_{uv}-\frac{1}{2}f(R)g_{uv}-\nabla_{u}
\nabla_{v}f_R+ g_{uv} \Box f_R=\kappa
(T_{uv}+E_{uv}),\quad(u, v =0,1,2,3),
\end{equation}
where $f_R\equiv df(R)/dR$, $\nabla_{u}$ denote covariant
derivative, $\Box=\nabla^{u}\nabla_{v}$, $T_{uv}$ is minimally
coupled stress-energy tensor and $E_{uv}$ is electromagnetic tensor.
Above field equations can also be written as
\begin{equation}\label{12}
G_{uv}=\frac{\kappa}{f_R}[L_{uv}],
\end{equation}
where $L_{uv}=\overset{(D)}{T_{uv}}+T_{uv}+E_{uv}$ with
\begin{equation}\label{d}
\overset{(D)}{T_{uv}}=\frac{1}{\kappa}\left[\frac{f(R)-Rf_R}{2}g_{uv}+\nabla_{u}
\nabla_{v}f_R -g_{uv} \Box f_R\right]
\end{equation}
denoting effective stress-energy tensor. The usual matter is
anisotropic and adiabatic in nature representing dissipative
collapse in the form of heat flux $q$ and is given by \cite{2, 20}
\begin{equation}\label{2}
T_{uv}=(\mu+p_\perp)V_{u}V_{v}-p_\perp g_{uv}+(p_r-p_\perp)\chi_u\chi_v+
q_u V_v+q_{v}V_u,
\end{equation}
where $\mu$ depicts density, $p_r$  to the radial pressure,
$p_\perp$ to the tangential pressure, $V_{u}$ to the four-velocity
of the fluid and $\chi_u$ corresponds to the radial four vector. In
co-moving coordinates, following pattern is accompanied
\begin{equation}\label{3}
V^{u}=A^{-1}\delta^{u}_{0},\quad q^{u}=qB^{-1}\delta^u_1,\quad
\chi^{u}\chi_{u}=-1, \quad \chi^{u}=B^{-1}\delta^{u}_{1}.
\end{equation}
The electromagnetic energy-momentum tensor is written as
\cite{Zaeem}
\begin{equation}\label{3'}
E_{uv}=\frac{1}{4\pi}(-F^w_{u}F_{vw}+\frac{1}{4\pi}F^{wx}F_{wx}g_{uv}).
\end{equation}
Here $F_{uv}=\varphi_{v,u}-\varphi_{u,v}$ denotes electromagnetic
field tensor while $\varphi_{u}=\varphi(t,r)\delta^{0}_{u}$ stands
for four potential. The Maxwell field equations are given by
\begin{equation}\label{3''}
F^{uv}_{;v}=\mu_0\jmath^u, \quad F_{uv;w}=0,
\end{equation}
where $\jmath^u=\mu(t,r)V^u$ is four current, $\mu_0$ is magnetic permeability and $\mu$ represents charge density.
The electromagnetic field equations turn out to be
\begin{eqnarray}
\label{M1}
&&\frac{\partial^2\varphi}{\partial r^2}-\left(\frac{A'}{A}+\frac{B'}{B}-\frac{2C'}{C}\right)
\frac{\partial\varphi}{\partial r}=4 \pi \mu AB^2,
\\\label{M2}&&\frac{\partial^2\varphi}{\partial t \partial r}-\left(\frac{\dot{A}}{A}
+\frac{\dot{B}}{B}-\frac{2\dot{C}}{C}\right)\frac{\partial\varphi}{\partial r}=0.
\end{eqnarray}
Herein derivatives with respect to $t$ and $r$ are expressed by dot and prime
respectively.
Applying integration on Eq.(\ref{M1}), we have
\begin{equation}\label{3''}
\frac{\partial\varphi}{\partial r}=\frac{qBA}{C^2}.
\end{equation}
The total charge $q$ interior to radius $r$ with electric field
intensity $E$ has the form
\begin{equation}\label{3''}
q={\int}^{r}_{0}\mu BC^2 dr,\quad E=\frac{q}{4\pi C^2}.
\end{equation}
For interior spacetime, the components on the right hand side of the
field equations (\ref{12}) are given as follows, whereas the
components of Einstein tensor are present in \cite{Riz4}
\begin{eqnarray}\nonumber
G_{00}&=&\frac{1}{f_R}\left\{\kappa\left(2\pi E^2+\rho\right)+ \frac{f-Rf_R}{2}+
\frac{f_R''}{B^2}-\frac{\dot f_R}{A^2}\left(\frac{\dot{B}}{B}+\frac{2\dot{C}}{C}\right)\right.\\\label{f1}
&&\left.
-\frac{f_R'}{B^2}\left(\frac{B'}{B}-\frac{2C'}{C}\right)\right\}
,\\\label{f2}
G_{01}&=&\frac{\kappa}{2f_R}\left\{qAB
+\frac{1}{\kappa}\left(\dot{f_R}'
-\frac{A'}{A}\dot{f_R}-\frac{\dot{B}}{B}f_R'\right)\right\},\\\nonumber
G_{11}&=&\frac{1}{f_R}\left[\kappa \left(p_r-2\pi E^2\right)- \frac{f-Rf_R}{2}+
\frac{\ddot{f_R}}{A^2}-\frac{\dot f_R}{A^2}\left(\frac{\dot{A}}{A}-\frac{2\dot{C}}{C}\right)
\right.\\\label{f3}
&&\left.-\frac{f_R'}{B^2}\left(\frac{B'}{B}+\frac{2C'}{C}\right)\right],
\\\nonumber
G_{22}&=&\frac{1}{f_R}\left[\kappa \left(p_\perp+2\pi E^2\right)-
\frac{f-Rf_R}{2}+
\frac{\ddot{f_R}}{A^2}-\frac{f_R''}{B^2}-\frac{\dot
f_R}{A^2}\left(\frac{\dot{A}}{A} \right.\right.\\\label{f4}
&&\left.\left.-\frac{\dot{B}}{B}-\frac{\dot{C}}{C}\right)
-\frac{f_R'}{B^2}\left(\frac{A'}{A}-\frac{B'}{B}+\frac{C'}{C}\right)\right].
\end{eqnarray}

The development in collapsing phenomenon with the passage of time
can be described by dynamical equations. These dynamical evolution
equations for usual matter, effective and Maxwell's energy-momentum
tensor carrying higher order curvature invariants are formed by
employing Bianchi identities as
\begin{eqnarray}\label{bb}
&&L^{uv}_{;v}V_{u}=0,\quad
L^{uv}_{;v}\chi_{u}=0,
\end{eqnarray}
which turn out to be
\begin{eqnarray}\nonumber &&\dot{\rho}+q'\frac{A}{B}+2q\frac{A}{B}\left(\frac{A'}{A}
+\frac{C'}{C}\right)+\rho\left(\frac{\dot{B}}{B}+\frac{\dot{C}}{C}\right)+
p_r\frac{\dot{B}}{B}+2p_\perp \frac{\dot{C}}{C}\\\label{B1}&&+4\pi
E^2\left(\frac{\dot{E}}{E}+ \frac{\dot{C}}{C}\right)+ P_1(r,t)=0,
\\\nonumber &&
p'_r+p_r\left(\frac{A'}{A}+\frac{C'}{C}\right)+\rho\frac{A'}{A}-2p_\perp\frac{C'}{C}+\dot{q}\frac{B}{A}+2\frac{B}{A}\left(\frac{\dot{B}}{B}
+\frac{\dot{C}}{C}\right)\\\label{B2}&&-4\pi
E^2\left(\frac{E'}{E}+\frac{2C'}{C}\right)+P_2(r,t)=0.
\end{eqnarray}
Here $P_1(r,t)$ and $P_2(r,t)$ corresponds to dark source terms provided in \textbf{Appendix} in the form of Eqs.(\ref{B3})
and (\ref{B4}) respectively.

\section{$f(R)$ Model and Perturbation Scheme}

The $f(R)$ model under consideration is
\begin{equation}\setcounter{equation}{1}\label{fr}
f(R)=R+{\alpha}{R^n}.
\end{equation}
The configuration of second order derivative decides whether the
model is viable or not. Any $f(R)$ is supposed to be suitable in GR
and Newtonian limits if $f''(R)>0$. For our proposed form of $f(R)$,
$n>2$ and $\alpha$ is a positive real number, in order to fulfill
stability criterion and demonstrate accelerated expansion of the
universe.

It is a fact that general solution of gravitational field equations
is yet unknown because these are highly complicated non-linear
differential equations. Perturbation theory can be employed to
somehow incorporate these discrepancies, so that dynamical equations
become linear in the form of metric and material variables.
Evolution can be explored by using Eulerian or Lagrangian schemes,
i.e., by using fixed or co-moving coordinates respectively. We have
applied perturbation with the assumption that initially all the
metric and material functions are in static equilibrium and with the
passage of time perturbed quantities have both radial and time
dependence. Carrying $0<\varepsilon\ll1$, functions may be written
in the following pattern

\begin{eqnarray}\label{41} A(t,r)&=&A_0(r)+\varepsilon
T(t)a(r),\\\label{42} B(t,r)&=&B_0(r)+\varepsilon T(t)b(r),\\\label{43}
C(t,r)&=&rB(t,r)[1+\varepsilon T(t)\bar{c}(r)],\\\label{44}
\rho(t,r)&=&\rho_0(r)+\varepsilon {\bar{\rho}}(t,r),\\\label{45}
p_r(t,r)&=&p_{r0}(r)+\varepsilon {\bar{p_r}}(t,r),
\\\label{46}
p_\perp(t,r)&=&p_{\perp0}(r)+\varepsilon {\bar{p_\perp}}(t,r), \\\label{47}
m(t,r)&=&m_0(r)+\varepsilon {\bar{m}}(t,r),\\\label{48}
q(t,r)&=&\varepsilon {\bar{q}}(t,r), \\\label{49'}
R(t,r)&=&R_0(r)+\varepsilon T(t)e(r),\\\label{50'}
E(t,r)&=&E_0(r)+\varepsilon T(t)h(r)\\\label{51'}
f(R)&=&[R_0(r)(1+\alpha R_0^{n-1}(r))]+\varepsilon
T(t)e(r)[1+\alpha n R_0^{n-1}(r)],\\\label{52'}
f_R(R)&=&1+\alpha n R_0^{n-1}(r)+\varepsilon\alpha n(n-1)R_0^{n-2}(r) T(t)e(r).
\end{eqnarray}
Assuming $C_0(r)=r$ as Schwarzschild coordinate, static
configuration of the field equations (\ref{f1})-(\ref{f4}) take the
following form
\begin{eqnarray}\nonumber
&&\frac{2B_0'}{rB_0}+\frac{B_0^2 -1}{r^2}=\frac{\kappa B_0^2}{1+\alpha n R_0^{n-1}}\left[\rho_0
+2\pi E_0^2+\frac{\alpha n(n-1)R_0^{n-2}}{\kappa}\left\{-\frac{R_0^{2}}{2n}\right.\right.\\\label{S1}&&\left.\left.
+\frac{(n-2)R_0^{-1}}{B_0^2}-\frac{1}{B_0^2}\left(\frac{B_0'}{B_0}-\frac{2}{r}\right)\right\}\right],
\end{eqnarray}
\begin{eqnarray}\nonumber
&&\frac{2A_0'}{rA_0}+\frac{B_0^2 +1}{r^2}=\frac{\kappa B_0^2}{1+\alpha n R_0^{n-1}}\left[p_{r0}-
2\pi E_0^2+\frac{\alpha n(n-1)R_0^{n-2}}{\kappa}\left\{\frac{R_0^{2}}{2n}\right.\right.\\\label{S2}&&\left.\left.
-\frac{1}{B_0^2}\left(\frac{A_0'}{A_0}-\frac{2}{r}\right)\right\}\right],
\\\nonumber
&&\frac{1}{r}\left(\frac{A_0'}{A_0}-\frac{B_0'}{B_0}\right)+\frac{A_0''}{A_0}-
\frac{A_0'B_0'}{A_0B_0}=\frac{\kappa B_0^2}{1+\alpha n R_0^{n-1}}\left[
\frac{\alpha n(n-1)R_0^{n-2}}{\kappa}\left\{\frac{R_0}{2n}\right.\right.\\\label{S3}&&\left.\left.
-\frac{(n-2)R_0^{-1}}{B_0^2}-\frac{1}{B_0^2}\left(\frac{A_0'}{A_0}-\frac{B_0'}{B_0}
+\frac{1}{r}\right)\right\}+p_{\perp0}+2\pi E_0^2\right].
\end{eqnarray}
First dynamical equation (\ref{B1}) is identically satisfied in static configuration,
while second evolution equation (\ref{B2}) has static configuration as under
\begin{eqnarray}\label{S4}
p_r'+(\rho_0+p_{r0})\frac{A_0'}{A_0}+(p_{r0}-p_{\perp0})\frac{2}{r}-4\pi E_0^2\left(\frac{E_0'}{E_0}+\frac{2}{r}\right)+P_{2s}=0,
\end{eqnarray}
where $P_{2s}$ depicts static part of $P_2(r,t)$ and provided in
\textbf{Appendix} as Eq.(\ref{S5}). Perturbed configuration of
Evolution equations (\ref{B1}) and (\ref{B2}) read
\begin{eqnarray}\nonumber &&
\dot{\bar{\rho}}-\bar{q}'\frac{A_0}{B_0}+2\bar{q}\frac{A_0}{B_0}
\left(\frac{A_0'}{A_0}+\frac{1}{r}\right)+ 2\pi E_0^2\left(\frac{h}{E_0}
+\frac{2\bar{c}}{r}\right)\\\label{B1p}&&+\dot{T}\left[\frac{b}{B_0}(\rho_0+p_{r0})
+\frac{2\bar{c}}{r}(\rho_0+p_{\perp0})+P_{1p}\right]=0
\\\nonumber
&&\bar{p_r}'+\dot{\bar{q}}\frac{B_0}{A_0}+(\bar{\rho}+\bar{p_r})\frac{A_0'}{A_0}
+(2\bar{p_r}+\bar{p_\perp})\frac{1}{r}
\\\nonumber
&&+\left[-4\pi\left\{(E_0 h )'+2E_0^2(\frac{\bar{c}}{r})'+2 E_0h(\frac{E_0'}{E_0}+\frac{2}{r})\right\}
\right.\\\label{B2p}
&&\left.+(\rho_0+p_{r0})(\frac{a}{A_0})'+(2p_{r0}+p_{\perp0})(\frac{\bar{c}}{r})'\right]T+P_{2p}
=0,
\end{eqnarray}
where $P_{1p}$ and $P_{2p}$ denote perturbed part of $P_{1}$ and
$P_{2}$ respectively and given in \textbf{Appendix}. Elimination of
$\bar{q}$ from perturbed equation (\ref{f2}) implies
\begin{eqnarray}\nonumber
\bar{q}&=&\frac{1}{\kappa
A_0B_0}\left[\alpha n(n-1)R_0^{n-2}\left\{e'+e(n-2)R_0^{-1}R_0'-e\frac{A_0'}{A_0}-\frac{b}{B_0}R_0'\right\}\right.\\\label{q}
&-&\left.2(1+\alpha n R_0^{n-1})\left\{\frac{\bar{c}A_0'}{rA_0}+\frac{b}{rB_0}-\frac{\bar{c}'}{r}\right\}\right]\dot{T}.
\end{eqnarray}
On substitution of $\bar{q}$ and its radial derivative in Eq.(\ref{B1p}) lead to an equation from which
$\dot{\bar{\rho}}$ can be extracted. Integrating this $\dot{\bar{\rho}}$ with respect to ``t", we get
\begin{eqnarray}\nonumber
\bar{\rho}&=&\left[-\frac{b}{B_0}(\rho_0+p_{r0})-\frac{2\bar{c}}{r}(\rho_0+p_{\perp0})
\right.\\\label{B6}&-&\left.4\pi E_0^2\left(\frac{h}{E_0}+\frac{2\bar{c}}{r}\right)+P_3(r)\right]T,
\end{eqnarray}
where $P_3(r)$ is presented in \textbf{Appendix}.
By second law of thermodynamics, $\bar{\rho}$ and $\bar{p_r}$ can be
related as ratio of specific heat with assumption of Harrison-Wheeler type
equation of state, expressed in following expression
\cite{ch1,HW}
\begin{equation}\label{B7}
\bar{p}_r=\Gamma\frac{p_{r0}}{\rho_0+p_{r0}}\bar{\rho}.
\end{equation}
The adiabatic index $\Gamma$ is a measure to recognize pressure
variation with changing density. Putting Eq.(\ref{B6}) in above
equation, we arrive at

\begin{eqnarray}\nonumber
\bar{p}_r&=&-\Gamma\left[p_{r0}\frac{b}{B_0}+\frac{2\bar{c}}{r}\frac{p_{r0}(\rho_0+p_{\perp0})}{\rho_0+p_{r0}}
\right.\\\label{B8}&+&\left.4\pi E_0^2\left(\frac{h}{E_0}
+\frac{2\bar{c}}{r}\right)\frac{p_{r0}}{\rho_0+p_{r0}}-\frac{p_{r0}}{\rho_0+p_{r0}}P_3\right]T.
\end{eqnarray}
Insertion of $\bar{\rho}$, $\bar{q}$ and $\bar{p}_r$ from
Eqs.(\ref{q}), (\ref{B6}) and (\ref{B8}) respectively in (\ref{B2p})
leads to
\begin{eqnarray}\nonumber&&
\frac{\ddot{T}}{\kappa
A_0^2}\left[\alpha n(n-1)R_0^{n-2}\left\{e'+e(n-2)R_0^{-1}R_0'-e\frac{A_0'}{A_0}-\frac{b}{B_0}R_0'\right\}-2(1\right.\\\nonumber
&&\left.+\alpha n R_0^{n-1})\left\{\frac{\bar{c}A_0'}{rA_0}+\frac{b}{rB_0}-\frac{\bar{c}'}{r}\right\}\right]
-\Gamma T\left[p_{r0}\frac{b}{B_0}+\frac{2\bar{c}}{r}\frac{p_{r0}(\rho_0+p_{\perp0})}{\rho_0+p_{r0}}\right.\\\nonumber&&\left.
+\left\{4\pi E_0^2\left(\frac{h}{E_0}
+\frac{2\bar{c}}{r}\right)-P_3\right\}\frac{p_{r0}}{\rho_0+p_{r0}}\right]_{,1}-
\Gamma T\left(\frac{A_0'}{A_0}+\frac{2}{r}\right)\left[p_{r0}\frac{b}{B_0}\right.\\\nonumber&&\left.
+\frac{2\bar{c}}{r}\frac{p_{r0}(\rho_0+p_{\perp0})}{\rho_0+p_{r0}}
+\left\{4\pi E_0^2\left(\frac{h}{E_0}
+\frac{2\bar{c}}{r}\right)-P_3\right\}\frac{p_{r0}}{\rho_0+p_{r0}}\right]+\frac{\bar{p}_\perp}{r}
\\\nonumber&&-\frac{A_0'}{A_0}\left[-\frac{b}{B_0}(\rho_0+p_{r0})-\frac{2\bar{c}}{r}(\rho_0+p_{\perp0})
-4\pi E_0^2\left(\frac{h}{E_0}+\frac{2\bar{c}}{r}\right)+P_3(r)\right]T\\\nonumber&&
+\left[-4\pi\left\{(E_0 h )'+2E_0^2(\frac{\bar{c}}{r})'+2 E_0h(\frac{E_0'}{E_0}+\frac{2}{r})\right\}
+(\rho_0+p_{r0})(\frac{a}{A_0})'\right.\\\label{B9}
&&\left.+(2p_{r0}+p_{\perp0})(\frac{\bar{c}}{r})'\right]T+P_{2p}=0.
\end{eqnarray}
When we perturb Ricci scalar curvature, we obtain the following
differential equation
\begin{equation}\label{66}
\ddot{T}(t)-P_4(r) T(t)=0.
\end{equation}
$P_4(r)$ is given in \textbf{Appendix}. The terms of equation $P_4$
are presumed in a way that it preserve positivity for establishment
of instability range. Consequently, the solution of Eq.(\ref{66}) is
obtained as
\begin{equation}\label{68}
T(t)=-e^{\sqrt{P_4}t}.
\end{equation}
To estimate instability range in Newtonian and post-Newtonian
limits, above equation would be used in Eq.(\ref{B9}).

\subsection*{Newtonian Regime}

In this approximation, we assume that $\rho_0\gg p_{r0}$, $\rho_0\gg
p_{\perp0}$ and $A_0=1,~B_0=1$. By substituting these values  in
Eq.(\ref{B9}), we find
\begin{eqnarray}\nonumber&&
\frac{\ddot{T}}{\kappa}\left[\alpha
n(n-1)R_0^{n-2}\{e'+e(n-2)R_0^{-1}R_0'-bR_0'\} -\frac{2(1+\alpha n
R_0^{n-1})}{r}[b-\bar{c}']\right]\\\nonumber &&-\Gamma
T\left[p_{r0}b+\frac{2\bar{c}}{r}p_{r0}\right]'- \Gamma
T\frac{2}{r}\left[p_{r0}b
+\frac{2\bar{c}}{r}p_{r0}\right]+\frac{\bar{p}_{\perp(N)}}{r}
+\left[(2p_{r0}+p_{\perp0})(\frac{\bar{c}}{r})'
\right.\\\label{B10}&&\left. -4\pi\left\{(E_0 h
)'+2E_0^2(\frac{\bar{c}}{r})'+2
E_0h(\frac{E_0'}{E_0}+\frac{2}{r})\right\}
+\rho_0a'\right]T+P_{2p(N)}=0.
\end{eqnarray}
Here $P_{2p(N)}$ denotes the Newtonian regime terms of perturbed
second Bianchi identity. Inserting value of $T$ from Eq.(\ref{68})
in the above equation and rearranging, we have
\begin{eqnarray}\label{74}
\Gamma<\frac{\rho_0a'-4\pi\left\{(E_0 h )'+2E_0^2(\frac{\bar{c}}{r})'+2E_0h(\frac{E_0'}{E_0}+\frac{2}{r})\right\}+P_{2p(N)}+P_{5}}
{\left[p_{r0}b+\frac{2\bar{c}}{r}p_{r0}\right]'+\frac{2p_{r0}}{r}\left[b
+\frac{2\bar{c}}{r}\right]},
\end{eqnarray}
where $P_5(r)$ is given in \textbf{Appendix}. It is worth mentioning here that
adiabatic index depends upon the electric field intensity, pressure components,
energy density and scalar curvature terms in this limit.
Thus, collapsing star would be unstable as long as inequality (\ref{74}) holds.

\subsubsection*{Asymptotic Behavior}

The expression for $\Gamma$ takes following form when $\alpha\rightarrow0$
\begin{eqnarray}\label{74as}
\Gamma<\frac{a'\rho_0-4\pi\left\{(E_0 h )'+2E_0^2(\frac{\bar{c}}{r})'+2E_0h(\frac{E_0'}{E_0}+\frac{2}{r})\right\}
-\frac{2P_4}{r\kappa}[b-\bar{c}']}
{\left[p_{r0}b+\frac{2\bar{c}}{r}p_{r0}\right]'+\frac{2p_{r0}}{r}\left[b
+\frac{2\bar{c}}{r}\right]}.
\end{eqnarray}
This result represents the Einstein solution.

\subsection*{Post Newtonian Regime}

Here, we analyze relativistic impressions upto $O(\frac{m_0}{r}+
\frac{Q^2}{2r^2})$. In this approximation, we take
\begin{eqnarray}\label{pn1}
&&A_0=1-\frac{m_0}{r}+\frac{Q^2}{2r^2},~~~~B_0=1+\frac{m_0}{r}-\frac{Q^2}{2r^2}, \\\label{pn2}
\Rightarrow~&&\frac{A_0'}{A_0}=\frac{2}{r}\frac{Q^2-rm_0}{2rm_0-2r^2-Q^2},~~~\frac{B_0'}{B_0}=\frac{2}{r}\frac{Q^2-rm_0}{2rm_0+2r^2-Q^2}.
\end{eqnarray}
Insertion of Eq.(\ref{pn1}) and (\ref{pn2}) in (\ref{B9}), $\Gamma$
reads
\begin{eqnarray}\label{PN}
&&\Gamma<\frac{W+X+\frac{\bar{p}_{\perp(PN)}}{r}+P_{2(PN)}}{N'-\frac{2}{r}\frac{rm_0-2r^2}{2rm_0-2r^2-Q^2}N},
\end{eqnarray}
where $P_{4(PN)},~ \bar{p}_{\perp(PN)}$ and $P_{2(PN)}$ corresponds
to PN regime terms of $P_4, \bar{p}_{\perp}$ and $P_2$ respectively.
However, $W$ and $X$ constitute the following expressions
\begin{eqnarray}\nonumber
W&=&\frac{4r^4P_{4(PN)}}{\kappa(2r^2+2rm_0-Q^2)^2}\left[ \alpha
n(n-1)R_0^{n-2}\{e'+e(n-2)R_0^{-1}R_0'\}\right.\\\nonumber&&\left.
-\frac{2}{r}\frac{Q^2-rm_0}{2rm_0-2r^2-Q^2} \left\{e\alpha
n(n-1)R_0^{n-2}+\frac{2\bar{c}}{r}(1+\alpha n R_0^{n-1})\right\}
\right.\\\nonumber&&\left.-\frac{2r^2}{2r^2+2rm_0-Q^2}\left\{b\alpha
n(n-1)R_0^{n-2}R_0' +\frac{2b}{r}(1+\alpha n R_0^{n-1})\right\}
\right.\\\label{pn3}&&\left. +\frac{2\bar{c}'}{r}(1+\alpha n
R_0^{n-1})\right],\\\nonumber
X&=&-\frac{2}{r}\frac{Q^2-rm_0}{2rm_0-2r^2-Q^2}\left[\frac{2br^2(\rho_0+p_r)}{2rm_0+2r^2-Q^2}
-\frac{2\bar{c}}{r}(\rho_0+p_\perp)+P_{4(PN)}
\right.\\\nonumber&-&\left. 4\pi
E_0^2\left(\frac{h}{E_0}+\frac{2\bar{c}}{r}\right)\right]+4\pi\left\{(E_0
h )'
+2E_0^2(\frac{\bar{c}}{r})'+2E_0h(\frac{E_0'}{E_0}+\frac{2}{r})\right\}\\\label{pn4}&-&
(\rho_0+p_r)\left(\frac{2ar^2}{2rm_0-2r^2+Q^2}\right)'+(2p_r+p_\perp)\left(\frac{\bar{c}}{r}\right)',
\\\nonumber
N&=&\frac{2br^2p_r}{2rm_0+2r^2-Q^2}
+\frac{p_r}{\rho_0+p_r}[4\pi E_0^2\left(\frac{h}{E_0}+\frac{2\bar{c}}{r}\right)-P_{4(PN)}]
\\\label{pn5}&+&\frac{2\bar{c}}{r}\frac{p_r(\rho_0+p_\perp)}{\rho_0+p_r}.
\end{eqnarray}
As far as instability problem concerned, system is unstable in PN
limit for the above inequality. We can see that how curvature terms
alongwith material variables affect the instability range. All terms
appeared in the above inequality must maintained positivity to
fulfill the dynamical instability condition and hence the following
constraints
\begin{eqnarray}\nonumber
N'>\frac{2}{r}\frac{rm_0-2r^2}{2rm_0-2r^2-Q^2}N, \quad
3rm_0>2(r^2+Q^2), \quad r^2<Q^2-2rm_0.
\end{eqnarray}

\subsubsection*{Asymptotic Behavior}

As $\alpha\rightarrow0$, adiabatic index $\Gamma$ is unchanged whereas $W$
becomes
\begin{eqnarray}\nonumber
W&=&\frac{4r^4}{\kappa(2rm_0+2r^2-Q^2)^2(2rm_0-2r^2-Q^2)}\left[\frac{2\bar{c}}{r^2}(Q^2
-rm_0)\right.\\\label{pn6}&&\left.-4br+\frac{2\bar{c}'}{r}\right].
\end{eqnarray}
This represents the GR solution in PN regime.

\section{Summary and Discussion}

The purpose of the present work is to determine the electromagnetic
field impressions on the instability of spherically symmetric
collapsing compact object in $f(R)$ framework. In order to achieve
the goal, locally anisotropic matter experiencing dissipative
collapse has been considered. We employ Jordan frame to work out
instability problem and to modify EH action for modified gravity, we
consider $f(R)=R+\alpha R^n$ and Maxwell source so that attributes
of $f(R)$ model alongwith electromagnetic field can be investigated.
Einstein field equations are modified accordingly. Dynamical
equations are developed to study the evolution of non-static
spherical star with the help of perturbation approach.

Model under consideration provides a viable alternate to dark energy, it satisfies the condition
$f''(R)>0$ to carry out stellar stable configuration for matter dominated regime.
Dissipation in terms of heat flow plays an important role in dynamics of collapse, especially
electric charge and its distribution imply drastic effects on evolution and stellar structure.

Since solution of field equations are not ascertained yet, that is
why to discuss dynamics of celestial bodies perturbation scheme is
used. Perturbed form of second Bianchi describes the evolution of
the collapsing system and further used to establish the instability
range in terms of adiabatic index $\Gamma$. It is evident from the
results that $\Gamma$ has dependency on electric field intensity,
radiative effects, density and pressure configuration. Inclusion of
Maxwell source alongwith higher order curvature invariants imply
that the self-gravitating system becomes more stable in the presence
of electromagnetic field. Results are reduced to GR as
$\alpha\rightarrow0$.

Lastly, we compare our findings with the previous literature and
found that our results reduces to the work already done for various
constraints on $n$ and electromagnetic effects. Comparison is
elaborated in the following
\begin{itemize}
\item For $p_r=p_\perp, n=-1$ and $\alpha=\delta^4$ our results reduce to isotropic pressure
case \cite{Z2}.
\item In the absence of Maxwell source results correspond to the results presented in
  \cite{Riz10}.
\item When we take $n=2$ in our Model, the results supports the arguments in \cite{Z3}. Also, it is clear
that addition of Maxwell invariant describe more general expanding
universe with a wider range of instability in the $f(R)$ framework.
\end{itemize}

\section{Appendix}

\begin{eqnarray}\setcounter{equation}{1}\nonumber
P_1(r,t)&=&\frac{1}{\kappa}\left[A^2\left\{\frac{1}{A^2}\left(\frac{f-Rf_R}{2}
-\frac{\dot{f_R}}{A^2}\left(\frac{\dot{B}}{B}+\frac{2\dot{C}}{C}\right)-\frac{f'_R}{B^2}\left(\frac{B'}{B}
-\frac{2C'}{C}\right)\right.\right.\right.\\\nonumber &&\left.\left.\left.+\frac{f''_R}{B^2}\right)\right\}_{,0}+A^2\left\{\frac{1}{A^2B^2}
\left(\dot{f'_R}-\frac{A'}{A}\dot{f_R}
-\frac{\dot{B}}{B}f_R'\right)\right\}_{,1}-\frac{\dot{f_R}}{A^2}\left\{\left(\frac{3A'}{A}\right.\right.\right.\\\nonumber
&&\left.\left.\left.+\frac{B'}{B}+\frac{2C'}{C}\right)\frac{AA'}{B^2}
+\left(\frac{\dot{B}}{B}\right)^2+2\left(\frac{\dot{C}}{C}\right)^2
+\frac{3\dot{A}}{A}\left(\frac{\dot{B}}{B}+\frac{2\dot{C}}{C}\right)
\right\}\right.\\\nonumber
&&\left.+\frac{\dot{f'_R}}{B^2}\left(\frac{3A'}{A}
+\frac{B'}{B}+\frac{2C'}{C}\right)-
\frac{2f'_R}{B^2}\left\{\frac{A'}{A}\left(\frac{2\dot{B}}{B}+\frac{\dot{C}}{C}\right)+\frac{B'}{B}\left(\frac{\dot{A}}{A}
\right.\right.\right.\\\nonumber &&\left.\left.\left.+\frac{\dot{B}}{B}\right)-
\frac{C'}{C}\left(\frac{2\dot{A}}{A}-\frac{\dot{B}}{B}+\frac{\dot{C}}{C}\right)\right\}
+\frac{f_R''}{B^2}\left(\frac{2\dot{A}}{A}+\frac{\dot{B}}{B}\right)+\frac{\dot{A}}{A}(f\right.\\\label{B3}
&&\left.
-Rf_R)+\frac{\ddot{f_R}}{A^2}\left(\frac{\dot{B}}{B}+\frac{2\dot{C}}{C}\right)\right],
\\\nonumber
P_2(r,t)&=&\frac{1}{\kappa}\left[B^2\left\{\frac{1}{B^2}\left(\frac{Rf_R-f}{2}
-\frac{\dot{f_R}}{A^2}\left(\frac{\dot{A}}{A}-\frac{2\dot{C}}{C}\right)-\frac{f'_R}{B^2}\left(\frac{A'}{A}
+\frac{2C'}{C}\right)\right.\right.\right.
\\\nonumber &&\left.\left.\left.+\frac{\ddot{f_R}}{A^2}\right)\right\}_{,1}
+B^2\left\{\frac{1}{A^2B^2}\left(\dot{f_R}'-\frac{A'}{A}\dot{f_R}
-\frac{\dot{B}}{B}f_R'\right)\right\}_{,0}+\frac{A'}{A}\left\{\frac{\ddot{f_R}}{A^2}
\right.\right.
\\\nonumber
&&\left.\left.+\frac{f_R''}{B^2}-\frac{\dot{f_R}}{A^2}\left(\frac{\dot{A}}{A}+\frac{\dot{B}}{B}\right)-
\frac{f'_R}{B^2}\left(\frac{A'}{A} +\frac{B'}{B}\right)\right\}
+\frac{2B'}{B}\left\{\frac{Rf_R-f}{2}\right.\right.\\\nonumber
&&\left.\left.+\frac{\ddot{f_R}}{A^2}
-\frac{\dot{f_R}}{A^2}\left(\frac{\dot{A}}{A}-\frac{2\dot{C}}{C}\right)
-\frac{f'_R}{B^2}\left(\frac{A'}{A}+\frac{3C'}{C}\right)\right\}
+\frac{1}{A^2}\left(\frac{\dot{A}}{A}+\frac{3\dot{B}}{B}\right.\right.
\\\nonumber &&\left.\left.
+\frac{2\dot{C}}{C}\right)\left(\dot{f_R}'-\frac{A'}{A}\dot{f_R}
-\frac{\dot{B}}{B}f_R'\right)+\frac{2C'}{C}\left\{\frac{f''_R}{B^2}+\frac{\dot{f_R}}{A^2}\left(\frac{\dot{C}}{C}
-\frac{2\dot{B}}{B}\right)\right.\right.\\\label{B4} &&\left.\left.
-\frac{f'_R}{B^2}\frac{C'}{C}\right\}\right].
\end{eqnarray}
Static part of $P_2(r,t)$ is
\begin{eqnarray}\nonumber
P_{2s}&=&\frac{\alpha n(n-1)}{\kappa}\left[B_0^2\left\{\frac{R_0^{n-2}}{nB_0^2}
\left(\frac{R_0^2}{2}-\frac{nR_0'}{B_0^2}\left(\frac{A_0'}{A_0}+\frac{2}{r}\right)\right)\right\}_{,1}\right.\\\nonumber&&\left.
+\left(\frac{A_0'}{A_0}+\frac{2}{r}\right)\left(\frac{R_0''+(n-2)R_0^{-1}{R_0'}^2}{B_0^2}\right)+\frac{2B_0'R_0^2}{2nB_0}
\right.\\\label{S5} &&\left.\frac{R_0^{n-2}R_0'}{B_0^2}\left\{\frac{A_0'}{A_0}\left(\frac{A_0'}{A_0}+\frac{B_0'}{B_0}\right)+
+\frac{2B_0'}{B_0}\left(\frac{A_0'}{A_0}+\frac{3}{r}\right)+\frac{2}{r^2}\right\}\right].
\end{eqnarray}
Perturbed form of $P_{1}$ and $P_{2}$ reads
\begin{eqnarray}\nonumber
P_{1p}&=&\frac{e''+\alpha n(n-1)R_0^{n-2}}{\kappa B_0^2}\left[
(n-2)\left\{(2e'-e)R_0^{-1}R_0'+eR_0^{-1}R_0''\right.\right.\\\nonumber&&\left.\left.
+e(n-3)R_0^{-2}{R_0'}^2\right\}-\frac{R_0B_0^2}{2}-\frac{b}{B_0}(R_0''
+(n-2)R_0^{-1}{R_0'}^2)\right.\\\nonumber&&\left.+e'+R_0'
\{\frac{\bar{c}}{r^2}+\frac{b'}{B_0}
+\frac{2\bar{c}'}{r}-\frac{b}{B_0}\left(\frac{2A'_0}{A_0}+\frac{3B'_0}{B_0}+\frac{4}{r}-1\right)
\right.\\\nonumber&&\left.
-\frac{2\bar{c}}{r}\left(\frac{A'_0}{A_0}-\frac{2}{r}\right)\}+\left(e'-e\frac{A'_0}{A_0}
+(n-2)eR_0^{-1}R_0'\right)\left(\frac{3A'_0}{A_0}
\right.\right.\\\nonumber&&\left.\left.+\frac{B'_0}{B_0}+\frac{1}{r}\right)\right]
+\frac{\alpha n(n-1)}{\kappa}\frac{A_0^2}{B_0^2}\left[\frac{R_0^{n-2}}{A_0^2B_0^2}\left\{e'-e\frac{A'_0}{A_0}
-\frac{b}{B_0}R_0'\right.\right.\\\label{P1p} &&\left.\left.
+(n-2)eR_0^{-1}R_0'\right\}\right]_{,1}
\\\nonumber
P_{2p}&=&\frac{\alpha n(n-1)}{\kappa}\left[\frac{R_0^{n-2}}{A_0^2}\ddot{T}\left\{(e'
+(n-2)eR_0^{-1}R_0')(1+A_0^2)-\frac{b}{B_0}R_0'
\right.\right.\\\nonumber&&\left.\left.
+2e(1-A_0^2)\frac{B'_0}{B_0}\right\}+2TbB_0\left\{\frac{R_0^{n-2}}{B_0^4}\{R_0'\left(\frac{A'_0}{A_0}+\frac{2}{r}\right)
-\frac{R_0^2B_0^2}{2n}\}\right\}_{,1}\right.
\\\nonumber&&\left.
+TB_0^2\left\{\frac{R_0^{n-2}}{B_0^4}\left[e\frac{R_0B_0^2}{2}+\left(\frac{A'_0}{A_0}+\frac{2}{r}\right)
\left\{e(n-2)R_0^{-1}R_0'+\frac{R_0^2bB_0}{n}
\right.\right.\right.\right.\\\nonumber&&\left.\left.\left.\left.
-R_0'[(\frac{a}{A_0})'+2(\frac{\bar{c}}{r})'+\frac{4b}{B_0}]+e'\right\}\right]\right\}_{,1}+
\frac{R_0^{n-2}}{B_0^2}T\left\{\left(\frac{A'_0}{A_0}+\frac{2}{r}\right)\{e''\right.\right.\\\nonumber&&\left.\left.
+[2e'R_0^{-1}R_0'+eR_0^{-1}R_0''
+e(n-3)R_0^{-2}{R_0'}^2](n-2)\}
-\left\{2\left(\frac{a}{A_0}\right)'\right.\right.\right.
\\\nonumber&&\left.\left.\left.
\times\left(\frac{\bar{c}}{r}\right)'+\frac{2b}{B_0}\left(\frac{A'_0}{A_0}
+\frac{2}{r}\right)\right\}[R_0''
+(n-2)R_0^{-1}{R_0'}^2]
+R_0'\left\{\left(\frac{a}{A_0}\right)'
\right.\right.\right.\end{eqnarray}
\begin{eqnarray}\nonumber&&\left.\left.\left.\times\left(\frac{2A'_0}{A_0}+\frac{3B'_0}{B_0}\right)
+3\left(\frac{b}{B_0}\right)'\left(\frac{A'_0}{A_0}+\frac{2}{r}\right)
+2\left(\frac{\bar{c}}{r}\right)'\left(\frac{3B_0'}{B_0}+\frac{2}{r}\right)\right\}
\right.\right.\\\nonumber&&\left.\left.
+[e'+e(n-2)R_0^{-1}R_0'-\frac{2b}{B_0}R_0'][\frac{A'_0}{A_0}\left(\frac{A'_0}{A_0}
+\frac{B'_0}{B_0}\right)+\frac{2}{r}\left(\frac{3B_0'}{B_0}\right.\right.\right.\\\label{P2p}&&\left.\left.\left.
+\frac{1}{r}\right)]\right\}\right]=0
\end{eqnarray}
Expression for $P_3$ and $P_4$ respectively is
\begin{eqnarray}\nonumber
P_{3}&=&-\frac{A_0}{ B_0}\left[\frac{1}{\kappa
A_0B_0}\left\{\alpha n(n-1)R_0^{n-2}\left(e'+e(n-2)R_0^{-1}R_0'-e\frac{A_0'}{A_0}
\right.\right.\right.\\\nonumber
&&\left.\left.\left.-\frac{b}{B_0}R_0'\right)-2(1+\alpha n R_0^{n-1})\left(\frac{\bar{c}A_0'}{rA_0}+\frac{b}{rB_0}-\frac{\bar{c}'}{r}\right)\right\}\right]_{,1}-
\frac{2}{\kappa B_0^2}
\\\nonumber
&&
\times\left[\alpha n(n-1)R_0^{n-2}\left(e'+e(n-2)R_0^{-1}R_0'-e\frac{A_0'}{A_0}-\frac{b}{B_0}R_0'\right)\right.\\\label{B5}&&\left.
-2(1+\alpha n R_0^{n-1})\left(\frac{\bar{c}A_0'}{rA_0}
+\frac{b}{rB_0}-
\frac{\bar{c}'}{r}\right)\right]\left(\frac{A'_0}{A_0}+\frac{1}{r}\right)-P_{1p}
\\\nonumber
P_{4}&=&-\frac{rA_0^2B_0}{br+2B_0\bar{c}}\left[ \frac{e}{2}-\frac{2\bar{c}}{r^3}
-\frac{1}{A_0B_0^2}\left\{A_0''[\frac{a}{A_0}+\frac{2b}{B_0}]-\frac{1}{B_0}\left(a'B_0'
\right.\right.\right.\\\nonumber && \left.\left.\left.
+a''+A_0'b'-A_0'B_0'[\frac{a}{A_0}+\frac{3b}{B_0}]\right)
+\frac{2}{r}\left\{a'+\bar{c}'A_0'-A_0'[\frac{a}{A_0}
\right.\right.\right.\\\nonumber && \left.\left.\left.+\frac{2b}{B_0}+\frac{\bar{c}}{r}]\right\}
+\frac{A_0}{r}\left\{\bar{c}''-\frac{b'}{B_0}-\frac{B_0'\bar{c}'}{B_0}+
\frac{3b}{B_0}+\frac{\bar{c}}{r}\right\}
+\frac{2}{r^2}[\bar{c}'\right.\right.\\\label{rp} && \left.\left.-\frac{b}{B_0}-\frac{\bar{c}}{r}]\right\}\right]
=0.\\\nonumber
P_{5}&=&\frac{P_4}{\kappa}\left[\alpha n(n-1)R_0^{n-2}\{e'+e(n-2)R_0^{-1}-bR_0'\}
\right.\\\label{N}&-&\left.\frac{2(1+\alpha n R_0^{n-1})}{r}[b-\bar{c}']\right]
\end{eqnarray}

\end{document}